\documentclass[preprint,12pt]{elsarticle}

\usepackage{amssymb}
\usepackage{amsmath}
\usepackage{bm}
\usepackage{xcolor}

\usepackage{lineno}
\usepackage{float}
\usepackage{hyperref}
\usepackage{tikz}
\usetikzlibrary{spy} 
\usepackage{siunitx}

\begin{document}

\begin{frontmatter}



\title{Particle migration in constricted flows: Effects of size and shape}

\author[a]{Raquel Dapena-García\corref{cor1}} 
\ead{raquel.dapena.garcia@usc.es}
\cortext[cor1]{Corresponding author.}
\author[a]{Vicente Pérez-Muñuzuri} 
\affiliation[a]{organization={CRETUS, Group of Nonlinear Physics, University of Santiago de Compostela, Spain},
            city={Santiago de Compostela},
            postcode={15782}, 
            country={Spain}}

\begin{abstract}
Cardiovascular diseases are a leading cause of death globally. Among them, some are linked to stenosis, which is an abnormal narrowing of blood vessels, as well as other factors. Smart drug delivery systems based on micro- and nanoparticles are a promising method to offer non/minimal-invasive therapeutic mechanisms. Here we investigate the propensity of particles with different shapes and sizes to drift laterally (marginate) towards an occlusion area in a two-dimensional (2D) parallel plate laminar flow using the Lattice-Boltzmann method (LBM). To verify the outcomes on both sides of the stenosis, a probability of adhesion to the borders was calculated. Analysis was done on the impact of wall-shear stress on both sides of the stenosis. Our results show that rectangular particles migrate in larger amounts and earlier than circular ones.
\end{abstract}




\end{frontmatter}


\section{\label{sec:intro}Introduction}

Particle migration in flows refers to the cross-streamline motion and redistribution of suspended particles or droplets within a flowing fluid. This can occur in a variety of contexts, such as in rivers, pipelines, and even in the human body. Particle migration can be affected by several factors, including the size and shape of the particles \cite{Liu2019}, the properties of the fluid, and the presence of external forces such as gravity or magnetism. Understanding particle migration is important in a wide range of fields, including chemical engineering, environmental science, and medicine \cite{BAHIRAEI201690}.

There has been a great deal of research on particle migration in flows in recent years, and significant progress has been made in understanding the underlying mechanisms and developing predictive models \cite{frank_anderson_weeks_morris_2003}. One area of active research is in developing methods to control or manipulate particle migration in flows, such as using external fields or by altering the properties of the fluid or particles \cite{KUMAR20201,ABDORAHIMZADEH2023}. Another focus is on understanding the impact of particle migration on the transport of heat, mass, and momentum in a system, as well as on the overall stability and performance of the flow \cite{MOLERUS1995871,Pirouz_2011}. Additionally, much work has been carried out on developing experimental techniques to measure and characterize particle migration in flows. Notable methods from previous decades include the use of optical scanning devices coupled with electronic counting circuits \cite{segre1961radial}, and setting large reservoirs to pump red cell-plasma suspensions through capillaries of different sizes \cite{barbee1971fahraeus}. More recently, advanced techniques such as laser-based imaging or tracking systems \cite{Coutinho_2023} have emerged as powerful tools to study this phenomenon. Interaction of particles with solid and elastic obstacles in the flow may lead to changes in lateral migration. 

Nanoparticles, particularly spheres, are typically proposed as carriers for vascular-targeted drug delivery due to their relative ease of fabrication and their capacity to navigate the circulatory system with minimal risk of vessel occlusion \cite{Amani_2021}. However, injected targeted therapeutics have often led to suboptimal and adverse systemic effects due to several factors, including early lateral migration and an overall inability to reach the designated target \cite{Chen_2006}. Some investigations have shown that efficacy of carriers for drug delivery depends on their ability to marginate (localize and adhere) to the vessel wall \cite{FOROUZANDEHMEHR2022110511}. In this regard, several studies suggest that, for particles circulating in a flow, the strategy of using larger particles to marginate could be more effective than using small ones \cite{DECUZZI2010320}. In fact, particles with submicrometer diameters have a lower capacity to bind the endothelium possibly because they are entrapped within the main flow \cite{Huang2010}. Furthermore, recent research shows the benefits of using non-spherical particles that can favor margination and interaction with the endothelium \cite{sharma2010polymer}. Particle size, shape, core-shell structure and stiffness play important roles in influencing particle behavior in blood flow in terms of cell-particle and vessel-particle interactions, transport and lateral margination. Computational models \cite{Decuzzi2005} have suggested that spherical particles tend to follow streamlines, while non-spherical particles (discs, rods…) can suffer significant lateral drift, even in the absence of external forces, which favors migration \cite{Gentile2008,Basagaoglu2018,Decuzzi2009,DOSHI2010196}. 

Several types of shear stress-sensitive drug vehicles are under research, in which drug delivery would be triggered by shear stress changes \cite{Korin2012,Korin_2015,Rana2019}. Recently, there has been an increased interest in shear-activated particle clusters for the treatment of stenosis \cite{Carboni2018}. However, to the best of our knowledge, there are no drug-carriers guided to the vascular lesion (by the flow characteristics and the change in wall shear stress) which enable a localized internalization in the diseased cells/tissues while providing a sustained release of the therapeutic cargo.

In recent years, Lattice-Boltzmann methods (LBMs) \cite{mohamad2019} have evolved enough to compete with classical tools in solving complex fluid flows problems, and be able to simulate particulate flows. Rooted in the mesoscale, the LBM models the fluid as a collection of particle distribution functions, and simulates their evolution over time. In this work, the LBM combined with a particle tracking scheme was used to study the lateral migration of particles of different shapes and sizes in a 2D stenotic domain. 



\section{Numerical method}

To study the motion of a particle cluster in a 2D constricted channel, the Lattice-Boltzmann method (LBM) \cite{succi2001lattice,mohamad2019,kruger2017lattice} is adopted. The dynamics of the Newtonian fluid flow with an external force can be described by a single relaxation time via the Bhatnagar-Gross-Krook (BGK) equation \cite{PhysRev.94.511},
\begin{eqnarray}
    f_i({\bf{r}}+{\bf{e}}_i\Delta t, t+\Delta t)-f_i({\bf{r}},t) &=& \frac{\Delta t}{\tau} \left[ f_i^{eq}({\bf{r}},t)-f_i({\bf{r}},t)\right] \nonumber \\ 
    & & +\Delta t S_i
\label{eq:distribution}    
\end{eqnarray}
where $f_i(\bf{r},t)$ is the distribution function at position $\bf{r}$ and time $t$, $f_i^{eq}(\bf{r},t)$ is the equilibrium distribution function \cite{Qian_1992}, $\tau$ is the dimensionless
relaxation time, and $\Delta t$ is the time step. In the present study, the D2Q9 (2-dimensional and 9-velocity) model \cite{mohamad2019} is adopted, where the equilibrium distribution function is defined as,
\begin{equation}
	f_i^{eq} = \omega_i \rho\left[ 1+\frac{{\bf{c}}_i \cdot \bf{u}}{c_s^2} + \frac{\left({\bf{c}}_i \cdot \bf{u}\right)^2}{2c_s^4} - \frac{u^2}{2c_s^2}\right]
\end{equation}

Here $\rho$ and $\bf{u}$ are the fluid density and velocity, respectively; $\omega_i$ represents the weighting parameter of the model, where $\omega_0 = 4/9$, $\omega_{1-4}= 1/9$ and $\omega_{5-8} = 1/36$; ${\bf{c}}_i$ is the discrete velocity of the model, with
$c_0 = (0, 0)$, $c_{1-4} = c(\cos[(i - 1)\pi/2], \sin[(i - 1)\pi/2])$, $c_{5-8}=2c(\cos[(2i - 1)\pi/4], \sin[(2i - 1)\pi/4])$; $c = \Delta x/\Delta t$ is the lattice speed, which was assumed to be one since $\Delta x = \Delta t = 1$; $\Delta x$ is the lattice spacing; and $c_s = c/\sqrt{3}$ is the sound speed of the model. The last term in Eq.~(\ref{eq:distribution}) represents the source term, which takes into account the force density driving the fluid as well as the forces exerted by particles, for which we use the Guo forcing scheme \cite{PhysRevE.65.046308}. Based on this scheme, a body force density $\bf{F}$ is related to $F_i$ as,
\begin{eqnarray}
	S_i&=& \left(1-\frac{\Delta t}{2\tau}\right) F_i \label{eq:S_i}\\
	F_i &=& \omega_i \left(\frac{\bf{c_i}-\bf{u}}{c_s^2}+\frac{\bf{c_i} \cdot \bf{u}}{c_s^4}\bf{c_i}\right)\bf{F} \label{eq:Fi}
\end{eqnarray}

The local fluid density, $\rho$, and velocity, $\bf{u}$, at the lattice node are given by $\rho = \sum_i f_i$ and $\rho {\bf{u}} = \sum_i f_i {\bf{c_i}} + \Delta t {\bf{F}}/2$. $\bf{F}$ is the sum of the forces acting on the fluid. The LBM for a single-phase flow recovers the Navier-Stokes equation in the limit of small Knudsen number for incompressible fluids, with the fluid kinematic viscosity being $\nu = c_s^2 (\tau - 0.5)$ (in lattice units). The kinematic viscosity cannot be less than zero; therefore, the limitation $\tau > 0.5$ must be satisfied.


The particle–fluid interaction was implemented using a force-coupling lattice Boltzmann method, conceptually following the formulation of Safdari and Kim \cite{safdari2014lattice}. In this approach, the particle surface is discretized into uniformly distributed boundary nodes, whose velocities are prescribed by the rigid-body motion of the particle. At each boundary node, the surrounding fluid velocity is interpolated and compared to the particle’s  velocity to evaluate the hydrodynamic forces and torque contributions.

The particles trajectories and velocities can be calculated from a Lagrangian point of view by integrating,
\begin{eqnarray}
     m_p\frac{d{\bf{v}}_p}{dt} &=& \bf{F_D} + \bf{F_L} \\
     I_p\frac{d\Omega_p}{dt} &=& \bf{T}
\end{eqnarray}
where $m_p$ and $I_p$ denote the mass and moment of inertia of the $p$ particle, ${\bf{v}}_p$ and $\Omega_p$ are the translational and rotational velocities of each particle, $\bf{F_D}$ represents the drag force, $\bf{F_L}$ is the lift force, and $\bf{T}$ the particle torque. Note that the effect of forces like virtual mass, Basset history and Magnus effect can be considered negligible. Other forces (Brownian, gravity, etc) acting on the particles are not considered in this paper. Particles' boundary are discretized in $M$ nodes as seen in Fig.~\ref{fig:lattices}, whose velocities ${\bf{v}}_b$ are given by,
\begin{equation}
    {\bf{v}}_b = {\bf{v}}_p + \Omega_p\times({\bf{r}}_b - {\bf{r}}_p)
\end{equation}
with ${\bf{r}}_b$ and ${\bf{r}}_p$ being the boundary node and particle positions, respectively.

\begin{figure}[ht]
	\begin{center}
		\begin{tikzpicture}[scale=0.75,spy using outlines={rectangle, magnification=3, size=3cm, connect spies}]
			
			\def\gridSize{6}
			\def\center{3}
			\def\radius{2}
			\def\nRed{12}
			\def\rRed{1.93} 
			
			\foreach \x in {0,...,8} {
				\foreach \y in {0,...,\gridSize} {
					\ifnum\x<8
					\draw[dashed, gray] (\x, \y) -- (\x+1, \y);
					\fi
					\ifnum\y<\gridSize
					\draw[dashed, gray] (\x, \y) -- (\x, \y+1);
					\fi
					\filldraw[white] (\x, \y) circle (0.04); 
					\draw (\x, \y) circle (0.09); 
				}
			}
			
			\foreach \y in {0,...,\gridSize}{
			\draw[dashed, gray] (-1, \y) -- (0, \y);
			\draw[dashed, gray] (8, \y) -- (9, \y);
			}
			
			\foreach \x in {0,...,8}{
			\draw[dashed, gray] (\x, -1) -- (\x, 0);
			\draw[dashed, gray] (\x, 6) -- (\x, 7);
			}
			
			\draw[red!50, dotted] (3.25,3.5) circle (\rRed);
			\foreach \i in {0,...,\numexpr\nRed-1} {
				\pgfmathsetmacro{\angle}{360/\nRed*\i}
				\fill[red] ({3.25+\rRed*cos(\angle)}, {3.5+\rRed*sin(\angle)}) circle (0.09);
			}
			
			\pgfmathsetmacro{\angle}{0}
			\pgfmathsetmacro{\xpos}{3.55+\rRed*cos(\angle)}
			\pgfmathsetmacro{\ypos}{3.5+\rRed*sin(\angle)}
			\draw[gray, thick, fill=gray, fill opacity=0.4] (\xpos-0.7,\ypos-0.7) rectangle (\xpos+0.7,\ypos+0.7);
			
			
		\end{tikzpicture}
	\end{center}
	\caption{Representation of a single discretized circular particle (red dots) moving through the  fluid lattice (white dots). Each particle boundary node is assigned a velocity $\bm{v_b}$. The corresponding fluid velocity at the same location, $\bm{u_b}$, is obtained via a four–neighbor interpolation, illustrated by the gray rectangle.}
	\label{fig:lattices}
\end{figure}
 
Finally, the drag and lift forces and the torque are given as,
\begin{eqnarray}
    \bf{F_D} &=& C_D \sum_b ({\bf{u}}_b - {\bf{v}}_p)\\
    \bf{F_L} &=& C_L \sum_b \left(({\bf{u}}_b - {\bf{v}}_p) \times \left[{\bf{\nabla}}\times({\bf{u}}_b - {\bf{v}}_p)\right]\right) \label{eq:lift} \\
    \bf{T} &=& C_T \sum_b \left(({\bf{r}}_b-{\bf{r}}_p)\times ({\bf{u}}_b - {\bf{v}}_p)\right)
\end{eqnarray}
where $C_D$, $C_L$ and $C_T$ denote the drag, lift, and torque coefficients, respectively, and ${\bf{u}}_b={\bf{u}} ({\bf{r}}_b)$ is the fluid velocity at the boundary nodes. The summation runs over all $M$ boundary nodes.

For the lift equation~(\ref{eq:lift}) we considered the Saffman force \cite{Saffman_1965} with
$C_L = 0.5$, adopted as a fixed, order-of-magnitude node coefficient in the absence of a
validated correlation for arbitrary particle shapes (see below). The drag coefficient is taken as a
function of the particle Reynolds number, following the Schiller--Naumann correlation
\cite{schiller1933uber} (see also \cite{michaelides}). We
further assume that the torque coefficient equals the drag coefficient, i.e.\
$C_T = C_D$; this simplification is examined against the classical Jeffery-orbit
rotation of an elongated particle in simple shear in the Supplementary Material, where
the resulting rotational dynamics reproduces the qualitative behaviour of the
analytical solution (continuous tumbling, non-uniform angular velocity slowing as the
long axis aligns with the flow). We emphasize that $C_L$ in Eq.~(\ref{eq:lift}) is not the
analytical Saffman lift coefficient for an isolated sphere in unbounded shear, which
assumes small particle Reynolds number and spherical symmetry and does not generally
extend to finite, non-spherical particles. Rather, $C_L$ acts at the level of each
discretized boundary node: shape and finite-size effects emerge from the number and
arrangement of the $M$ nodes on the particle surface, whose individual contributions sum to give the particle's shape-dependent lift response. This reflects the fact that shear-induced lift for particles of arbitrary shape remains an open modelling question \cite{Ravnik_2013}, motivating our use of a uniform node coefficient in the absence of a shape-specific correlation for
arbitrary aspect ratio rectangles. Indirect support for this treatment is provided by
the single-particle benchmark reported in the Supplementary Material, where an
off-centre neutrally buoyant particle migrates toward the channel centreline,
consistent in sign with lift-driven inertial migration.

Then, the position and velocity of each particle is updated by a forward explicit Euler scheme:
\begin{eqnarray}
	{\bf{r}}_p(t+\Delta t) = {\bf{r}}_p(t) + \Delta t {\bf{v}}_p\\
	{\bf{v}}_p(t+\Delta t) = {\bf{v}}_p(t) + \Delta t {\bf{F}}
\end{eqnarray}
with $\bf{F}=\bf{F_D}+\bf{F_L}$. The accuracy of this coupling procedure, as well as the overall implementation, has been validated against standard benchmarks, which are reported in the Supplementary Material.

\begin{figure}[ht]
\centering
\includegraphics[width=0.8\linewidth]{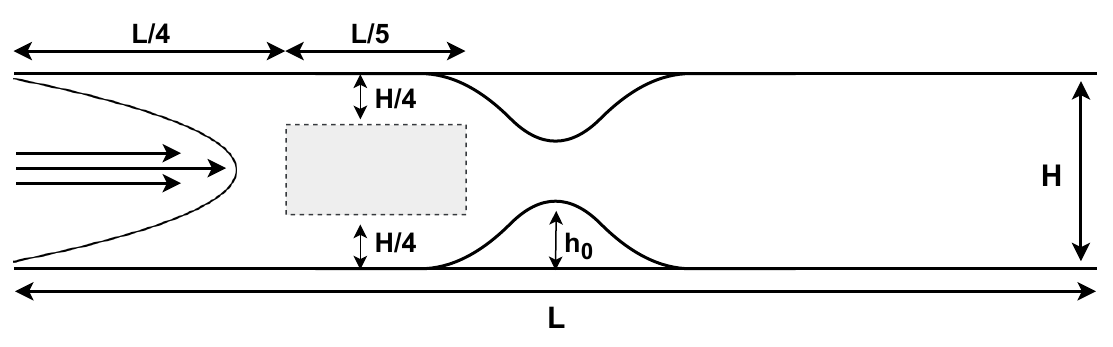}
\caption{Schematic of the 2D setup used in the simulations.}
\label{fig:setup}
\end{figure}


Figure~\ref{fig:setup} shows the 2D rectangular domain ($H\times L=400\times4000$ cells) considered (see~\ref{app:grid_convergence} for a grid convergence study). The constriction zone or stenosis is simulated as a Gaussian function of height $h_0$ and variance $h_0^2$ centered at $L/2$. A Poiseuille flow with a maximum velocity $u_{\max}$ at $H/2$ was considered at the inlet and an outflow boundary condition was imposed for the outlet, both achieved through the use of Zou/He type boundary conditions on the mesoscopic level \cite{zou1997pressure}. The bounce-back scheme was implemented on channel top and bottom walls---including the curved Gaussian regions---to simulate the no-slip condition on the fluid, capturing the wall and inertial effects on particle-fluid motion, as discussed by Basagaoglu et al. (2008) \cite{PhysRevE.77.031405}. 

Particles had initially the same velocity as the flow at the release location. In this paper, we focus specifically on circular and rectangular rigid particles. $N$ particles were released randomly far enough from the constriction zone and at $H/4$ cells far from the lateral boundaries (gray area in Fig.~\ref{fig:setup}). Care was put in making sure there is no overlapping between them. Rectangles were set initially at random orientations. To verify that the results were not affected by the random initialization, simulations were repeated using different random seeds, yielding consistent results. For a circular particle with radius $R$ rotating around its center of mass, the angular moment of inertia is $I_p=m_p R^2/2$, while for a rectangular particle with sides $l_1$ and $l_2$, $l_2\ge l_1$, $I_p=m_p(l_1^2+l_2^2)/12$. We consider elastic collisions between two particles, or a particle and a wall. For the collision between circular particles we suppose a hard sphere collision, whereas for the collision between two rectangles we follow Qin et al. (2020) \cite{QIN2020109021}. For the last case, the collision model takes into account whether the collision involves two vertices, a vertex and an edge or two edges. In this paper, we do not consider collisions between circular and rectangular particles. To compare results between both types of particles, we define the equivalent radius for the rectangle particles as $R_{equiv} = l_1(1+\gamma^2)^{1/2}/2$ where the aspect ratio is defined as $\gamma=l_2/l_1$. The plasma is modeled as a Newtonian fluid and initialized with a density $\rho=1$. The maximum flow velocity is set to $u_{\max}=0.04$, with a bulk Reynolds number $Re=H u_{\max}/\nu$ and a relaxation time $\tau=0.62$, all given in lattice units. Henceforth, all results presented in this work are expressed in lattice units. To facilitate the interpretation of the simulation results in a physiological context, the main physical parameters together with the corresponding conversion factors between lattice and physical units are summarized in Table~\ref{tab:units}. Note that the chosen parameters lie within the physiological range reported for blood flow in human carotid arteries \cite[Chapter~1]{sidebotham2007cardiothoracic}, ensuring that the simulated flow conditions are representative of real hemodynamic behaviour.

\begin{table}[H]
\centering
\begin{tabular}{|c|c|c|c|}
\hline
Quantity & Symbol & Value in l.u & Value in p.u \\ \hline
Vessel diameter & $H$ & 400 & \SI{4}{\milli\meter} \\
Plasma kinematic viscosity & $\nu$ & 0.04 & \SI{1.2e-6}{\meter\squared\per\second} \\
Plasma density & $\rho$ & 1  & \SI{1025}{\kilogram\per\meter\cubed} \\
Fluid peak velocity & $u_{\max}$ & 0.04  & \SI{0.12}{\meter\per\second} \\
Lattice spacing & $\Delta x$ & 1 & \SI{10}{\micro\meter} \\
Lattice time step & $\Delta t$ & 1 & \SI{3.33}{\micro\second} \\
\hline
\end{tabular}
\caption{Overview of the fundamental simulation parameters and equivalence between lattice units (l.u) and physical units (p.u).}
\label{tab:units}
\end{table}

\section{Results}

\subsection{\label{sec:margination}Flow trajectories of particles } 

\begin{figure}[ht]
\centering
\includegraphics[width=0.8\linewidth]{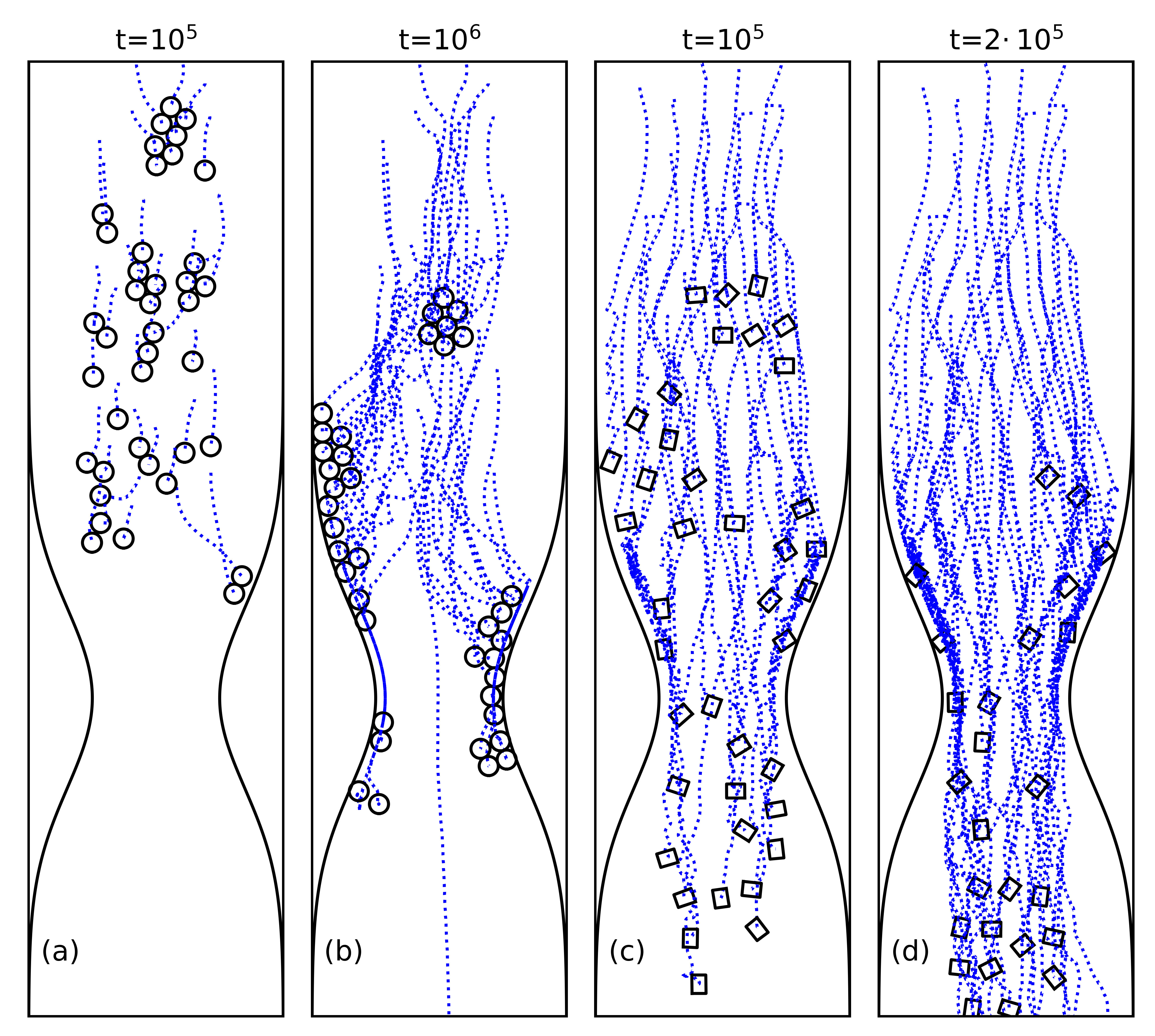}
\caption{Trajectories and particles positions for different shapes; (a-b) circular, $R=15$ and (c-d) rectangular, $R_{equiv}=18$ ($l_1=22, \gamma=1.3$) at different times crossing a $50\%$ constriction zone. In every case, the initial positions of the particles were the same. Set of parameters: $Re=400$ and $N=40$.}
\label{fig:part_pos}
\end{figure}

The LBM was used to simulate the migration trajectories of particles as a function of their size and shape. Simulation results in Fig.~\ref{fig:part_pos} unveiled two distinct shape dependent particle behaviors in a constricted channel: (i) both circular and rectangular particles migrate towards both lateral boundaries; however, the rectangles do so before the disks. Additionally, (ii) rectangular particles migrate to both sides of the stenosis and in greater quantity than disks. This finding aligns with previous studies focusing on the behaviour of particles with different aspect ratios ranging from spherical to discoidal \cite{gentile2008effect,Ye2021}. Particles migrate toward the lateral walls as they approach the stenosis, then pass through the opening, where they are entrained by the main jet and exit. Some particles remain grouped in small clusters that move more slowly than the particles that passed earlier. These aggregates reduce the fluid's velocity and modify the surrounding flow. Note that both types of particles create a wake, in which others are caught and therefore experience less drag, resulting in slower motion and
eventual clustering. This effect is weaker for rectangular particles, which exhibit cumulative angular rotation that enhances dispersion, unlike circular particles.

Particles roll along the walls of the geometry. Their probability of adhering to the boundaries will be considered later. One may find a small overlapping area of touching particles, especially for rectangles. This is expected from the numerical method because of the smooth transition of the object boundaries. The size of the overlapping area was reduced to its minimum size adjusting the relaxation time $\tau$.

To investigate the dispersion of particles with different aspect ratios, we calculate the mean square separation between particle pairs,
\begin{equation}
    MSS(t) = \big\langle \big(d_{ij}(t)-d_{ij}(0)\big)^2 \big\rangle_{i<j}, \qquad d_{ij}(t)=\|\bm{r}_i(t)-\bm{r}_j(t)\|
\label{eq:MSS}
\end{equation}
where $\bm{r}_k$ is the position of the $k$-th particle and the average is taken over all distinct particle pairs $i<j$. We further define the lateral (cross-stream) mean square displacement of individual particles relative to their initial position,
\begin{equation}
    MSD_{y,p}(t) = \big\langle \big|y_k(t)-y_k(0)\big| \big\rangle_k
\label{eq:MSDy}
\end{equation}
where $y_k$ denotes the lateral coordinate of particle $k$. Both quantities exhibit a power-law growth at intermediate times, $MSS(t)\propto t^{\alpha}$ and $MSD_{y,p}(t)\propto t^{\alpha_T}$, with the exponents $\alpha$ and $\alpha_T$ extracted from a log-log linear fit.

\begin{figure}[ht]
\centering
\includegraphics[width=0.95\linewidth]{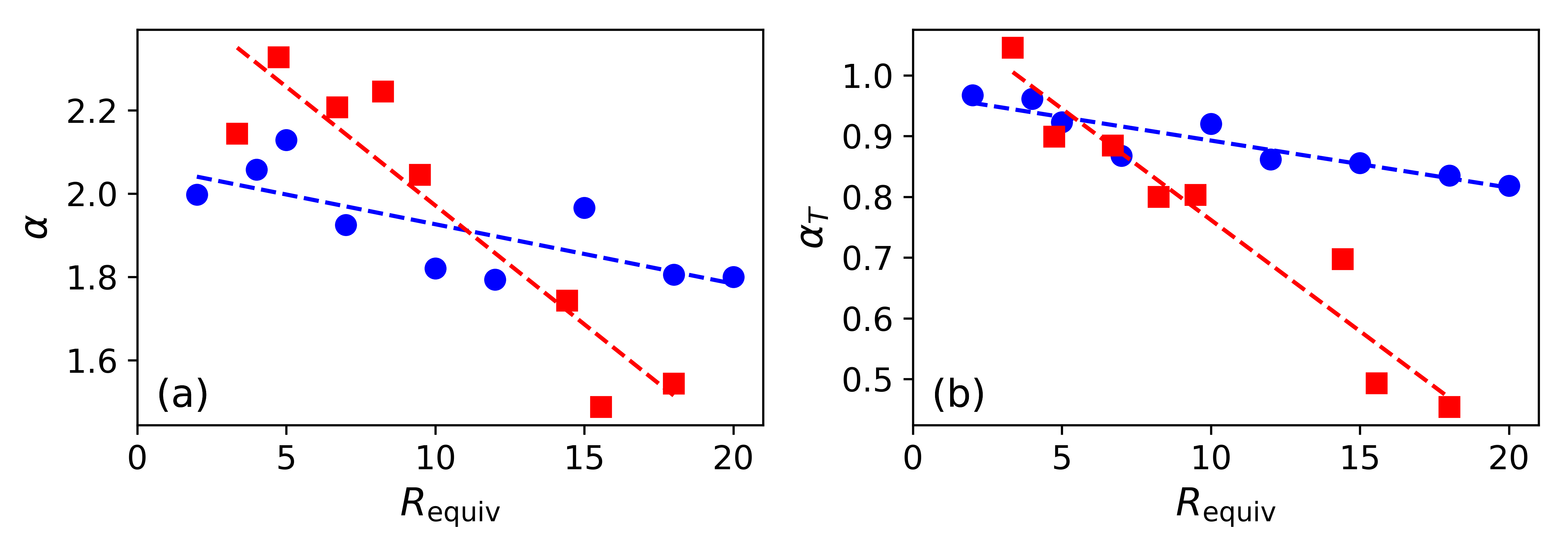}
\caption{Exponents of the power-law fits $MSS(t)\propto t^{\alpha}$ (a) and $MSD_{y,p}(t)\propto t^{\alpha_T}$ (b) as a function of the equivalent radius for disks (blue circles) and rectangles (red squares). Both exponents were calculated as the slope of a log-log fitting of $MSS(t)$ and $MSD_{y,p}(t)$, respectively. Dashed lines are shown to guide the eye. Set of parameters as in Fig.~\ref{fig:part_pos}.}
\label{fig:rate}
\end{figure}

Figure~\ref{fig:rate} shows $\alpha$ (a) and $\alpha_T$ (b) as a function of the
equivalent radius for disks and rectangles. Because $MSS(t)$ is built from the full 2D
pair separation $d_{ij}$ (Eq.~(\ref{eq:MSS})), it reflects both lateral and streamwise separation;
in a Poiseuille-type flow, particles at different lateral positions are advected at
different streamwise speeds, so $\alpha$ is best read as a measure of overall pairwise
dispersion rather than lateral migration on its own. The transverse displacement
$MSD_{y,p}(t)$ (Eq.~(\ref{eq:MSDy})), by construction, isolates the lateral coordinate and is
unaffected by this axial contribution; we therefore base our conclusions about lateral
migration on $\alpha_T$, and report $\alpha$ as a complementary, less specific
diagnostic of pair dispersion.

For disks, $\alpha_T$ stays close to unity ($\alpha_T \approx 0.85$--$0.95$) across the
whole size range, indicating that circular particles continue to drift laterally in a
persistent, near-directed manner irrespective of size. Similarly, $\alpha$ remains close to its
ballistic value ($\alpha \approx 2$) over the same range, consistent with--though not
uniquely attributable to--this persistent lateral drift, since a comparable value would
also be expected from streamwise shear dispersion alone. Rectangular particles, in
contrast, show $\alpha_T$ decreasing sharply with $R_{\mathrm{equiv}}$, from near unity
at small sizes down to strongly sub-diffusive values at large sizes ($\alpha_T \approx
0.45$--$0.5$). This shape-dependent saturation of $\alpha_T$ indicates that large
rectangular particles reach their equilibrium lateral position earlier and subsequently
oscillate around it, rather than continuing to migrate across streamlines, consistent
with the enhanced and earlier margination of rectangles shown earlier. The parallel
decay of $\alpha$ for rectangles is consistent with this picture. Once laterally
confined, large rectangles would also be expected to separate from one another more
slowly along the streamwise direction, likely due to increased correlated motion
(clustering/rolling along the boundary) among particles sharing a similar near-wall
trajectory, although we note that $\alpha$ alone cannot distinguish this from a purely
streamwise effect.


\begin{figure}[ht]
\centering
\includegraphics[width=0.8\linewidth]{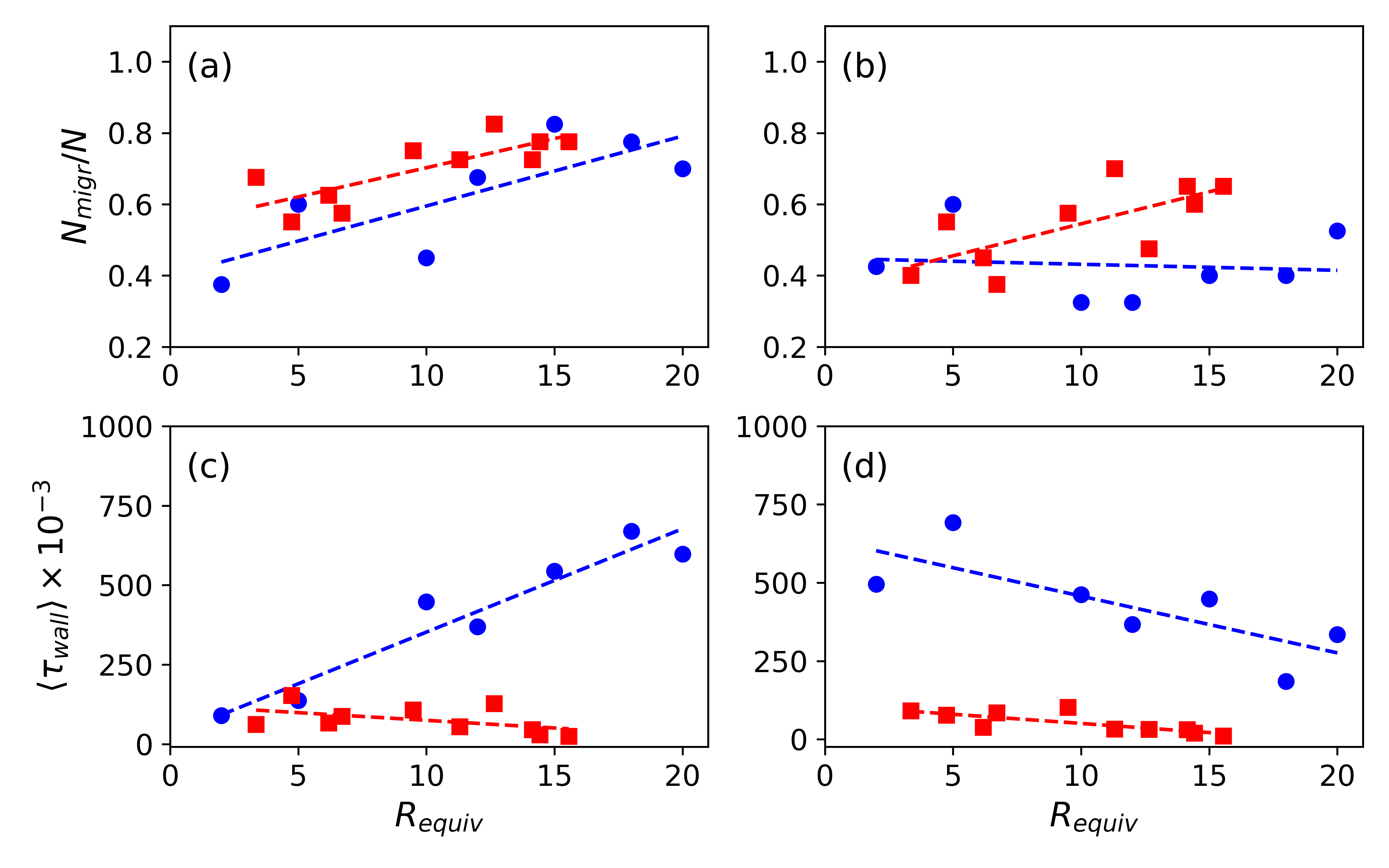}
\caption{Ratio of particles migrated to the lateral boundaries and average residence time at the boundaries prior to (a,c) and following (b,d) the stenosis as a function of $R_{\mathrm{equiv}}$. Disks and rectangular particles are represented by blue circles and red square dots, respectively. To guide the eye, dashed lines are displayed. Set of parameters as in Fig.~\ref{fig:part_pos}.}
\label{fig:part_Requiv}
\end{figure}

Particle migration to the edges depends on their size and shape as their response to drag and lift forces is different. If a particle node is located within the intervals $[L/2-5h_0,L/2]$ before and $[L/2,L/2+5h_0]$ after the stenosis, and if the distance between it and the boundaries is less than 5 cells, the particle is said to have migrated to the walls. Additionally, the average amount of time particles spend in these regions was computed. The ratio of migrated particles $N_{migr}/N$ and the wall residence time $\langle \tau_{\mathrm{wall}} \rangle$ for disks and rectangles as a function of $R_{\mathrm{equiv}}$ before (a,c) and after (b,d) the stenosis are displayed in Fig.~\ref{fig:part_Requiv}. Both magnitudes were computed for the same number of iterations. Following the stenosis, fewer particles (disks and rectangles) reach the channel boundaries compared to the upstream region. For the rectangular particles, $N_{migr}$ grows with $R_{\mathrm{equiv}}$ on both sides of the stenosis; however, no tendency was found for the disks downstream of the constriction. Although lift forces increase—enhancing the likelihood of rectangular particles migrating toward the walls—disks are dragged by the centre  flow, where the velocity is at its maximum. Small circular particles are confined in the channel's center region and rarely have their trajectories altered. 

The mean wall residence time $\langle \tau_{\mathrm{wall}} \rangle$ exhibits distinct upstream and downstream trends as a function of the equivalent radius $R_{\mathrm{equiv}}$.  Upstream of the stenosis (panel~c), $\langle \tau_{\mathrm{wall}} \rangle$ increases with particle size for disks, indicating enhanced wall interactions as larger particles search for admissible trajectories through the constriction. Downstream of the stenosis (panel~d), the opposite trend is observed: larger disks display shorter wall residence times. This reduction arises because larger particles exit the constriction centrally focused and experience stronger steric and hydrodynamic repulsion from the channel walls, which suppresses transverse diffusion toward the boundaries. In contrast, for rectangular particles, $\langle \tau_{\mathrm{wall}} \rangle$ decreases monotonically with increasing $R_{\mathrm{equiv}}$ both upstream and downstream, suggesting that shape-induced constraints dominate over size-dependent entropic trapping.

Very large particles can collapse/stop the flow at the stenosis by stacking \cite{Dressaire2017,Trofa_2021,Herale_2025}. It should be noted that in three-dimensional (3D) simulations, although particle accumulation may occur, the fluid velocity would not drop to zero, as there is space for the fluid to flow around the particles even if they partially obstruct the stenosis throat \cite{li2004lattice}. 

The role of the initial positioning of the particles was also investigated, although the results are not shown here. Particles injected far from the stenosis are more likely to be dragged by the central flow, with this phenomenon being more probable for circular particles than for rectangular ones.

\begin{figure}[ht]
\centering
\includegraphics[width=0.8\linewidth]{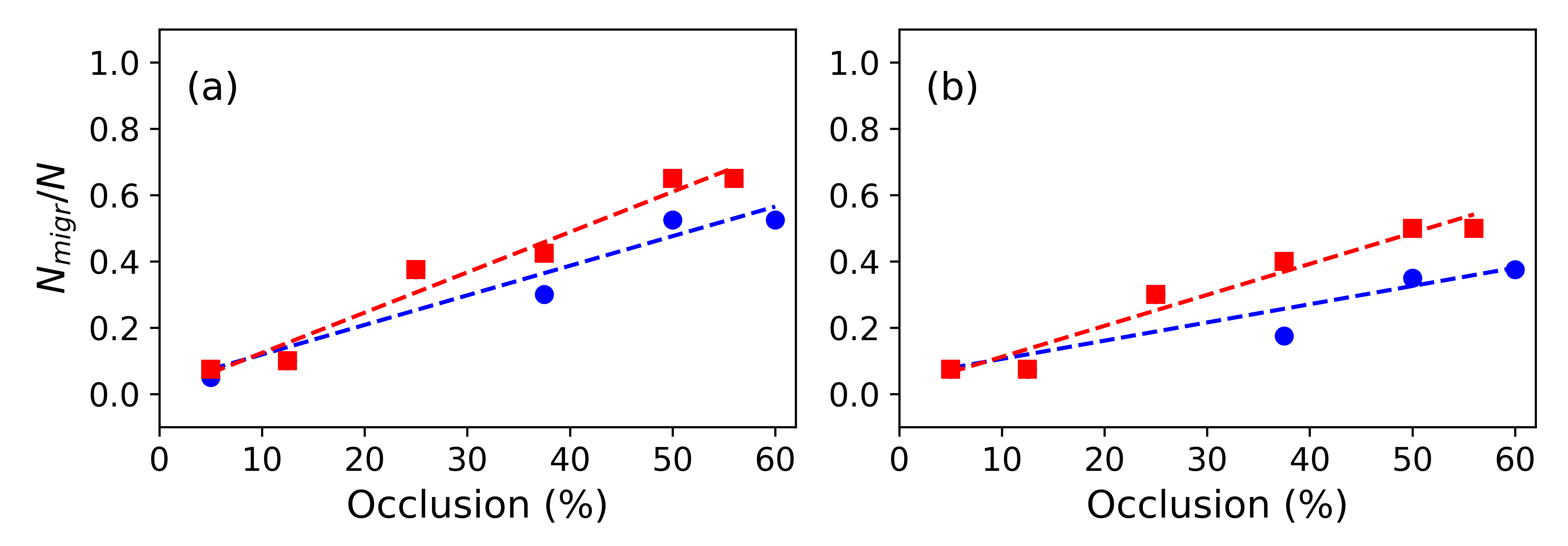}
\caption{Ratio of particles migrated to the lateral boundaries before (a) and after (b) the stenosis as a function of the occlusion percentage. Blue line corresponds to disks with $R=10$ and red line to rectangles with $R_{equiv}=9.5$ ($l_1=6, \gamma=3$). Set of parameters as in Fig.~\ref{fig:part_pos}.}
\label{fig:part_h0}
\end{figure}

The influence of the occlusion percentage on particle migration is shown in Fig.~\ref{fig:part_h0}. For both types of particles with similar values of $R_{\mathrm{equiv}}$, the number of particles that migrate to the boundaries increases with an increasing occlusion. Wall-shear stress upstream of the stenosis increases with $h_0$, favoring migration of particles to the boundaries; conversely, post-stenosis wall-shear stress diminishes due to the presence of a recirculation zone, thus reducing the rate of particle migration, especially for circular particles, which are more prone to be dragged by the flow. The driving forces for margination are significantly altered by the constriction.

\begin{figure}[ht]
\centering
\includegraphics[width=0.8\linewidth]{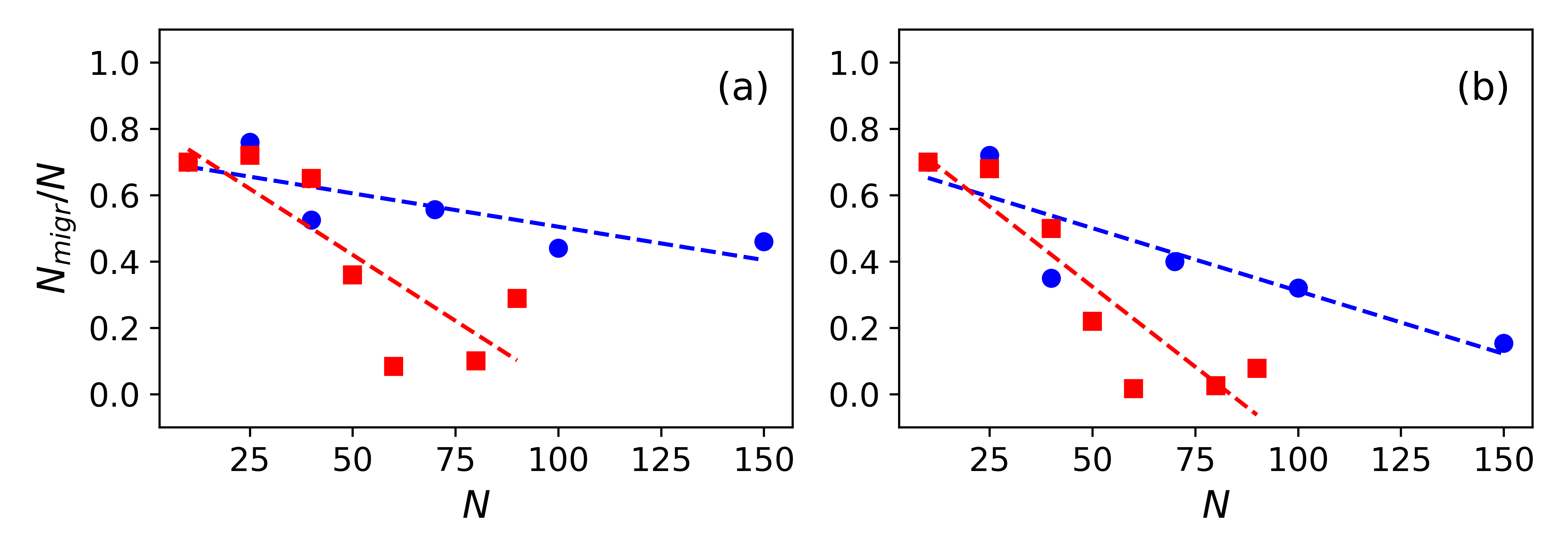}
\caption{Ratio of particles migrated to the lateral boundaries before (a) and after (b) the stenosis as a function of the number of released particles $N$. Blue line corresponds to disks with $R=10$ and red line to rectangles with $R_{equiv}=9.5$ ($l_1=6, \gamma=3$). Set of parameters as in Fig.~\ref{fig:part_pos}.}
\label{fig:part_Np}
\end{figure}

\begin{figure}[ht]
\centering
\includegraphics[width=0.7\linewidth]{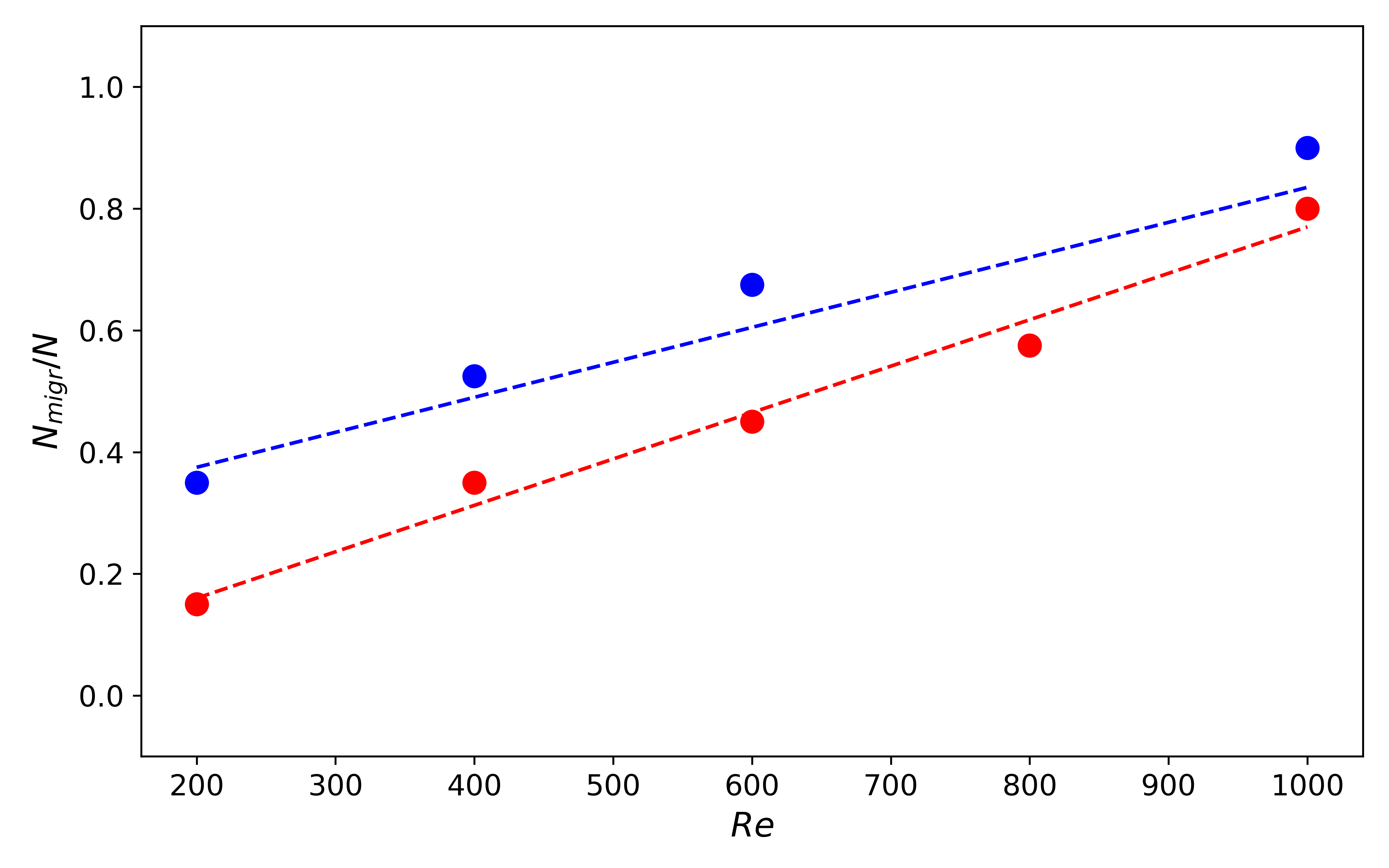}
\caption{Ratio of circular particles migrated to the lateral boundaries before (blue line) and after (red line) the stenosis as a function of the bulk Reynolds number. Set of parameters as in Fig.~\ref{fig:part_pos}.}
\label{fig:part_Re}
\end{figure}

Figure~\ref{fig:part_Np} illustrates the influence of the number of released circular particles $N$ on the ratio of particles that migrated to the boundaries. The results presented indicate that $N_{migr}/N$ decreases with $N$ before and after the stenosis, as increasing $N$ leads to larger slow flow at the particles wake and clogging at the constriction, thus reducing the ratio of particles that migrate to the lateral boundaries with respect to the number of emitted particles. This effect is most pronounced downstream of the constriction for circular particles, whose migration ratio declines substantially more steeply after the stenosis than before it; rectangular particles, in contrast, already show a comparably steep decline upstream of the constriction, which persists but is only modestly enhanced downstream. The effect of the bulk Reynolds number on the particle migration is analyzed in Fig.~\ref{fig:part_Re}. As with the other parameters, all Reynolds numbers considered lie within the physiologically relevant range for standard arteries. Only results containing circular particles are displayed; earlier results obtained for rectangular particles indicate that $N_{migr}$ is greater than that of circular particles. Increasing the Reynolds number by increasing $u_{\max}$ leads to a linear increase of $N_{migr}$, as the drag and lift forces are proportional to $Re$. 

\subsection{Probability of adhesion}

Particle migration occurs naturally due to flow conditions, being faster and in greater quantity for non circular shapes. However, particles were not intended to adhere to the boundaries so they may roll down the walls and thus modify the number of particles migrating to either side of the stenosis.

There are numerous models for calculating the general adhesion strength \cite{Bell_1984,Hammer_1987,Sun2003}. In this paper we used a simplified version of the bioadhesion model by Decuzzi and Ferrari \cite{DECUZZI20065307}, in which the adhesion strength directly correlates with the higher chance of successful bonds of particles to the boundaries, $P$. Once a bond is formed, adhesion is assumed to be irreversible, and the particle remains permanently attached to the vessel wall.

To use this model, we adopt the same threshold of 5 cells as in the margination calculation, beyond which adhesion is not considered possible. Thus, for particles of oval-like shape,
\begin{eqnarray}
    P &=& \pi r_0^2\exp\left[ -2.4\cdot 10^{-4} f R_{\mathrm{equiv}} \mu_s/r_0^2\right] \label{eq:P_adh}\\
    r_0^2 &=& R_{\mathrm{equiv}}^2 \left[1-\left(1-0.05\gamma/R_{\mathrm{equiv}}\right)^2 \right] \\
    f &=& 6\left(R_{\mathrm{equiv}}\gamma^{-1}+1\right)F^s + 8R_{\mathrm{equiv}}^2T^s/r_0
\end{eqnarray}
where $F^s$ and $T^s$ are coefficients that hinge upon the aspect ratio of particles $\gamma$, and $\mu_s$ is the wall-shear stress \cite{DECUZZI20065307}. Eq.~\ref{eq:P_adh} shows that, in this formulation, adhesion strength is primarily governed by the geometric properties of the particle, namely its radius $R_{\mathrm{equiv}}$ and the aspect ratio $\gamma$. 

\begin{figure}[ht]
\centering
\includegraphics[width=0.7\linewidth]{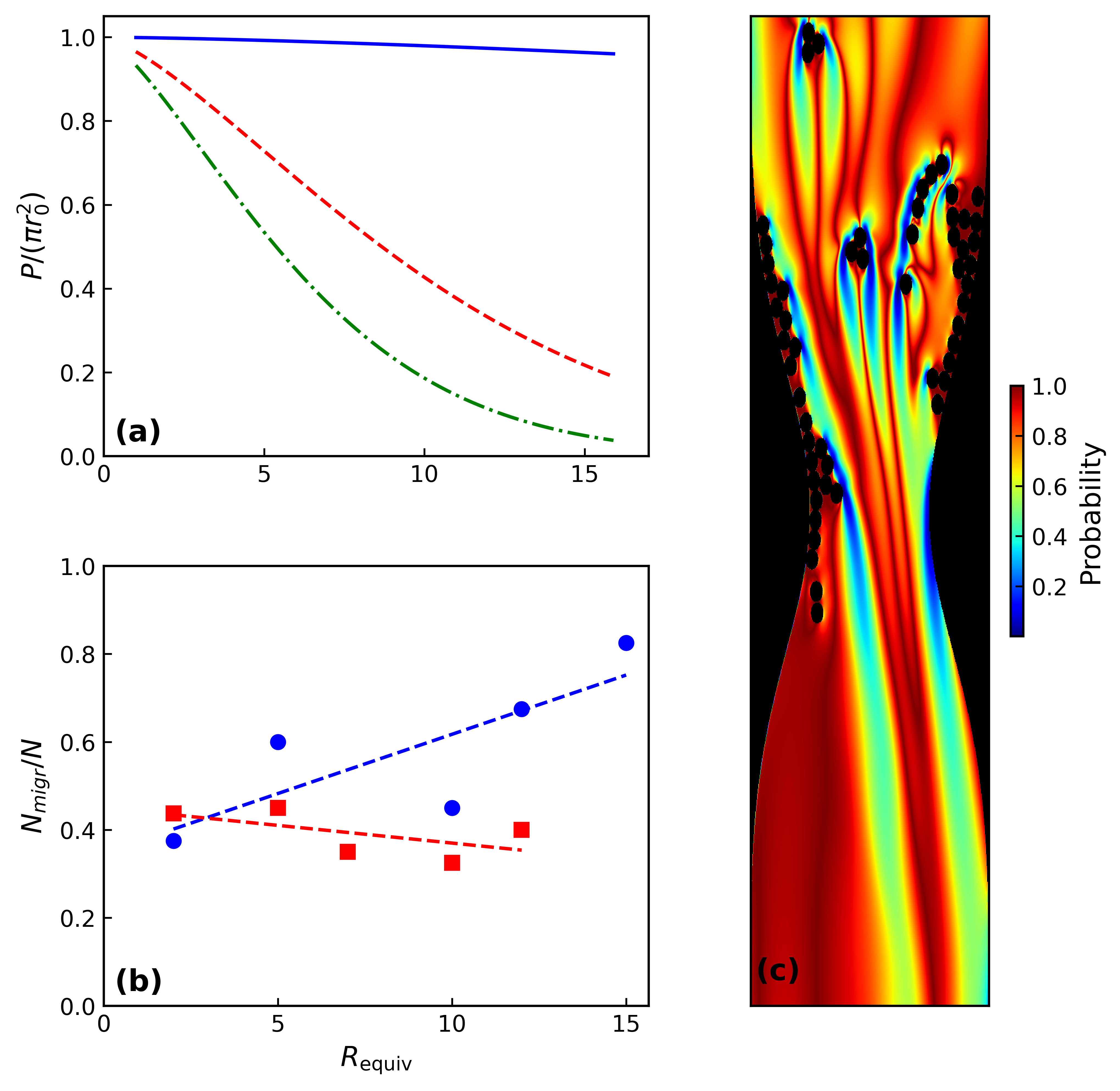}
\caption{Adhesive probability for three different wall shear stress values (a), and
ratio of migrated--without considering adhesion--(blue line) and adhered (red line)
circular particles before the stenosis (b) as a function of $R_{\mathrm{equiv}}$. Shear
stress increases from the upper line (blue) to the lowest line (green) in (a). Panel (c) displays the bonding probability for the entire flow. Particles in the flow have an
impact on the shear stress and, in turn, $P$. Set of parameters: $Re = 400$, $N = 80$
and percentage of occlusion 50\%.}
\label{fig:prob_adh}
\end{figure}

The variation of the normalized adhesion probability, $P/(\pi r_0^2)$, with the radius $R_{\mathrm{equiv}}$ of circular particles under three different wall shear stress conditions is shown in Fig.~\ref{fig:prob_adh}(a). The adhesion probability decreases with increasing shear stress and particle size. The three shear stress values correspond to distinct regions of the numerical domain: a shear acceleration zone at the proximal end of the stenosis (red dashed line), a peak shear zone immediately upstream of the stenosis apex (green dot-dashed line), and a shear deceleration zone at the distal end of the stenosis within the recirculation region (blue solid line). As expected, regions characterized by lower wall shear stress exhibit higher adhesion probabilities \cite{DECUZZI20065307}.

The presence of particles alters the local shear stress distribution and, consequently, the adhesion probability, as shown in panel (c). For example, in the low-bonding-probability region at the distal end of the constricted channel, particle-induced shear variations significantly affect adhesion behavior. 
All these trends are consistent with previous findings \cite{DECUZZI20065307}.

As discussed in the previous section, increasing particle radius promotes particle migration toward the vessel walls upstream of the stenosis (blue curve in Fig.~\ref{fig:prob_adh}(b)). However, despite this enhanced margination, the number of particles adhering to the walls decreases with increasing radius, consistent with the reduction in $P$ (red curve in panel~b). Downstream of the stenosis, fewer circular particles migrate toward and adhere to the walls, even though $P$ increases due to reduced shear stress. This behavior can be attributed to flow acceleration near the stenosis apex, which traps particles and hinders their approach to the vessel boundaries. Post-stenosis, results not depicted here exhibit a similar trend but with reduced values. 


Adhesion of particles to the lateral boundaries does not occur continuously in time. Instead, this behavior can be described statistically by a waiting-time distribution $q(\tau_w) \propto \tau_w^{-\beta}$, where the time intervals between two successive particle adhesion events are referred to as {\em waiting times} $\tau_w$. The estimated value of the exponent $\beta$ can be influenced by the finite duration of the integration time or by the limited number of particles, which may lead to an underestimation of long waiting times. Nevertheless, results indicate that the exponent $\beta \approx 0.77 \pm 0.11$ is negligibly affected by $R_{\mathrm{equiv}}$ \cite{DECUZZI20065307}.

\begin{figure}[ht]
\centering
\includegraphics[width=0.7\linewidth]{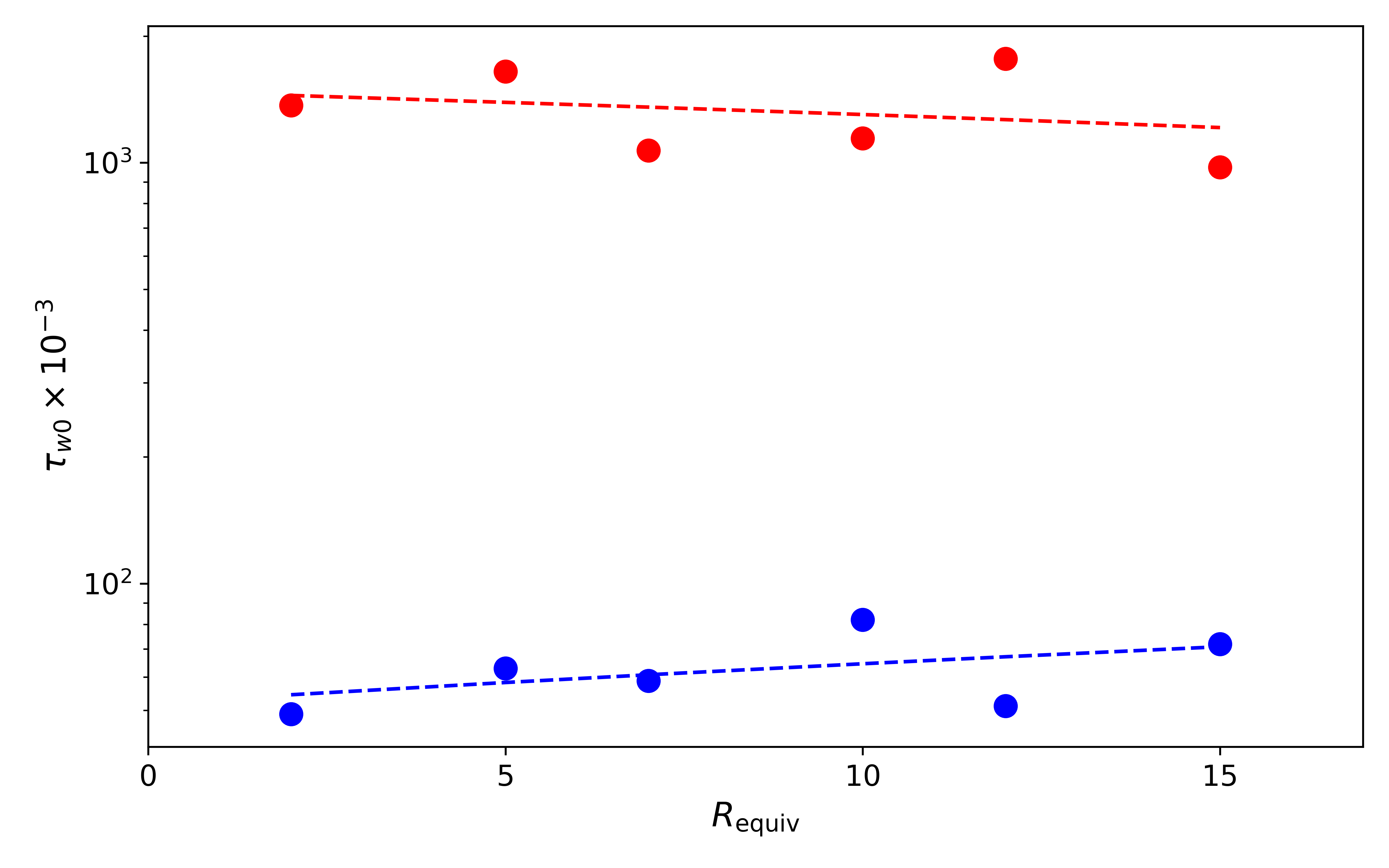}
\caption{Initial period of time for circular particles adhesion $\tau_{w_0}$ to the lateral boundaries as a function of $R_{\mathrm{equiv}}$ before (lower blue line) and after (upper red line) the stenosis. Dashed lines are shown as a visual guide. Set of parameters: $Re=400$, $N=80$ and percentage of occlusion 50\%.}
\label{fig:tw0_adh}
\end{figure}

Figure~\ref{fig:tw0_adh} shows the initial adhesion time $\tau_{w_0}$ of circular particles to the boundaries, both upstream and downstream of the stenosis. The observed behavior is similar to that found for the wall residence time in Fig.~\ref{fig:part_Requiv}, namely an increase (decrease) of $\tau_{w_0}$ upstream (downstream) of the stenosis as $R_{\mathrm{equiv}}$ increases. Before the stenosis apex, smaller disks adhere more rapidly than larger ones. In contrast, downstream of the constriction, smaller disks are more strongly dragged by the flow and therefore require a longer time to migrate toward the boundaries and subsequently adhere to them.

\section{Conclusions}

Particle--particle and particle--wall interactions in a constricted channel were investigated using the Lattice-Boltzmann method, with particular emphasis on the effects of particle size, shape, and adhesion probability to the boundaries.

Rectangular particles exhibited greater lateral migration toward the channel walls on both sides of the constricted region compared to circular particles. They also showed higher sensitivity to variations in fluid flow, responding more rapidly in terms of momentum. Nevertheless, particle size remained the dominant parameter governing transport behavior, with smaller particles reacting more quickly in all cases. Notably, the use of medium-sized rectangular particles, rather than large circular ones, resulted in enhanced particle entrapment on both sides of the stenosis. This finding suggests a strategy to improve the efficacy of drug-delivery carriers without relying on very large particles or high particle concentrations that could increase the risk of vessel occlusion.

Particle size was also found to influence the mean wall residence time. The residence time increased with particle radius upstream of the stenosis, but decreased with increasing disk size downstream. Due to the presence of recirculating flow in the wake of the stenosis, particle migration increased with occlusion severity, particularly for rectangular particles downstream of the constricted region.

Stronger adhesion promoted earlier particle migration toward the channel walls on both sides of the constricted zone. In contrast, upstream regions characterized by high wall shear stress exhibited a lower probability of particle adhesion. Local flow characteristics and particle position were found to modify the wall shear stress, thereby influencing the particle--wall bonding probability. Future studies should address the potential for particle aggregation and pile formation at the boundaries, which may lead to clustering effects capable of significantly altering the flow velocity within the constricted region. Moreover, incorporating a probabilistic description of particle detachment under varying shear conditions would enable a more realistic assessment of the balance between particle retention and release. Such an approach is essential for optimizing drug-delivery efficiency by ensuring sufficient residence time for therapeutic action while simultaneously reducing the risk of excessive accumulation and particle-induced clogging in stenotic vessels.

It should be noted that extending the present analysis to three dimensions could alter the quantitative outcomes. In such geometries, secondary flows, radial shear gradients, and out-of-plane particle motions can modify migration dynamics compared to the 2D case presented here. Blockage of blood flow in the constricted region due to particle clogging is unlikely in 3D vessels, reducing the likelihood of complete obstruction. In addition, particle orientation and contact area with the vessel wall are more complex in 3D, potentially influencing adhesion probabilities. While the qualitative trends reported here---such as the dependence of margination and adhesion on particle size and aspect ratio---are expected to remain valid, the precise migration rates and adhesion efficiencies may differ. A full three-dimensional study would therefore be an important direction for future work.

Overall, the results presented in this work provide physics-level insight relevant to carrier design considerations targeting stenotic regions.

\section*{CRediT authorship contribution statement}
\textbf{Raquel Dapena-García:} Methodology, Conceptualization, Validation, Formal analysis, Writing - Original Draft. \textbf{Vicente Pérez-Muñuzuri:} Conceptualization, Writing - Review $\&$ Editing, Supervision, Funding acquisition, Project administration. 

\section*{Declaration of competing interest}
The authors have no conflicts to disclose.

\section*{Data availability}
Source code for the LBM simulations is available from the authors upon reasonable request.

\section*{Acknowledgements}
This research has been supported by the Xunta de Galicia (Grant No. 2025-PG025) and Ministerio de Ciencia e Innovación (Grant No. PID2022-141626NB-I00).

\appendix

\section{\label{app:grid_convergence}Grid convergence details}

The numerical method was evaluated for four distinct grids ($H=200$, $400$, $600$, $800$ and $L=10\times H$) for circular particles moving through a stenotic two-dimensional channel with two percentages of occlusion values at $Re=400$ and $N=100$. To that end, the time-averaged particle Reynolds number was calculated:
\begin{equation}
<Re_p> =\frac{1}{N T}\sum_{t,k} \frac{2 R {\bf{v}}_k }{\nu}
\label{eq:Re_p}
\end{equation}
with ${\bf{v}}_k$ being the k-th particle velocity at time $t$. Particle radius ($R=5$, $10$, $15$, $20$), stenosis height $h_0$, and initial fluid velocity $u_{\max}$ were renormalized appropriately as $H\times L$ was varied, and to keep the Reynolds number $Re=400$ constant with $\nu=0.04$, which was widely used in the results shown above. The number of time steps $T$ in each simulation was increased by a factor of 4 to simulate the same physical time in each case. $<Re_p>$ calculated for the three larger grids are within $\approx 4\%$ error as shown in Table~\ref{tab:grid}, with the discrepancies being due to the random initial particles positions, and low number of particles. As mean particle velocities in the central flow increase with occlusion, $<Re_p>$ also does. Ultimately, in our simulations, to be computationally efficient, we have chosen the medium grid ($H\times L=400\times4000$ cells).  

\begin{table}[H]
\centering
\begin{tabular}{|c|c|c|}
\hline
\textbf{$H \times   L$} & \textbf{$<Re_p>   (25\%)$} & \textbf{$<Re_p>   (50\%)$} \\ \hline
$200 \times   2000$     &           0.21                &           0.28                 \\ \hline
$400 \times   4000$     &           0.12                &           0.14                 \\ \hline
$600 \times   6000$     &           0.11                &           0.13                 \\ \hline
$800 \times   8000$     &           0.12                &           0.13                 \\ \hline
\end{tabular}
\caption{Mean particle Reynolds number for two occlusion values (25\% and 50\%) and four grid sizes $H\times L$.}
\label{tab:grid}
\end{table}

\bibliographystyle{elsarticle-num} 
\bibliography{bib_margination.bib}

\end{document}